\documentclass[journal]{IEEEtran}
\usepackage{graphicx,subfigure,amsmath,amsfonts,amssymb,booktabs,cite,multirow,color}

\begin{document}

\title{In Vitro Fertilization (IVF) Cumulative Pregnancy Rate Prediction from Basic Patient Characteristics}

\author{Bo~Zhang, Yuqi~Cui, Meng~Wang, Jingjing~Li, Lei~Jin and Dongrui~Wu,
\thanks{B.~Zhang, M.~Wang, J.~Li and L.~Jin are with the Reproductive Medicine Center, Tongji Hospital, Tongji Medical College, Huazhong University of Science and Technology, Wuhan, Hubei 430030, China. Email: dramanda@126.com, tjmu\_mu\_wm@163.com, 1402964400@qq.com, jinleitjh@126.com.}
\thanks{Y.~Cui and D.~Wu are with the Key Laboratory of the Ministry of Education for Image Processing and Intelligent Control, School of Artificial Intelligence and Automation, Huazhong University of Science and Technology, Wuhan, Hubei 430074, China. Email: Yuqicui@hust.edu.cn, drwu@hust.edu.cn.}

\thanks{The first two authors contributed equally to this work.}
\thanks{Corresponding authors: Lei~Jin (jinleitjh@126.com), Dongrui~Wu (drwu@hust.edu.cn).}}

\maketitle

\begin{abstract}
Tens of millions of women suffer from infertility worldwide each year. In vitro fertilization  (IVF) is the best choice for many such patients. However, IVF is expensive, time-consuming, and both physically and emotionally demanding. The first question that a patient usually asks before the IVF is how likely she will conceive, given her basic medical examination information. This paper proposes three approaches to predict the cumulative pregnancy rate after multiple oocyte pickup cycles. Experiments on 11,190 patients showed that first clustering the patients into different groups and then building a support vector machine model for each group can achieve the best overall performance. Our model could be a quick and economic approach for reliably estimating the cumulative pregnancy rate for a patient, given only her basic medical examination information, well before starting the actual IVF procedure. The predictions can help the patient make optimal decisions on whether to use her own oocyte or donor oocyte, how many oocyte pickup cycles she may need, whether to use embryo frozen, etc. They will also reduce the patient's cost and time to pregnancy, and improve her quality of life.
\end{abstract}


\IEEEpeerreviewmaketitle

\section{Introduction}

According to the World Health Organization (WHO) \cite{Zegers-Hochschild2009}, infertility is ``\emph{a disease of the reproductive system defined by the failure to achieve a clinical pregnancy after 12 months or more of regular unprotected sexual intercourse.}" For women under 60, infertility was ranked the 5th highest serious global disability \cite{WHO2011}. Estimates from 25 international population surveys sampling 172,413 women indicated that 9\% of them suffered from infertility \cite{Boivin2007}. Another study \cite{Mascarenhas2012} on household survey data from 277 demographic and reproductive health surveys for women aged 20-44 estimated that 48.5 million couples worldwide suffered from infertility in 2010. The 2006-2010 United States National Survey of Family Growth (NSFG)  \cite{Chandra2013} sampling 22,682 men and women aged 15-44 also found that 6.0\% (1.5 million) women suffered from infertility in 2006-2010.

Assisted reproductive technology (ART) \cite{Sunderam2017} could help these couples to conceive pregnancy. The most common ART is in vitro fertilization (IVF) \cite{Chen2018}, which retrieves eggs from a woman's ovaries, fertilizes them in the laboratory, and then transfers the resulting embryos into the woman's uterus through the cervix. According to the 2015 ART National Summary Report \cite{ART2015}, more than 99\% ART cycles performed in the United States in 2015 used IVF.

\begin{figure*} \centering
\includegraphics[width=.9\linewidth]{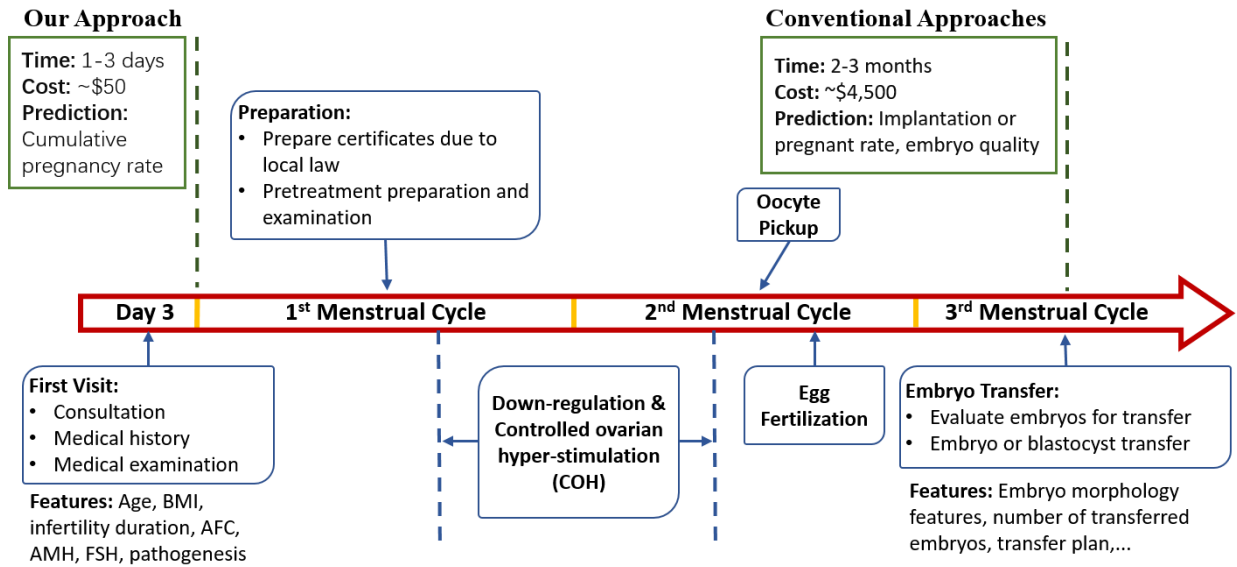}
\caption{The IVF timeline. Our model utilizes only the basic medical examination information during the first visit, and it can give the cumulative pregnancy rate prediction on Day~3 when the initial medical examination results are ready. Conventional approaches in the literature use information during the actual IVF to predict the pregnancy rate, and hence are much more time-consuming and expensive than our approach.}
\label{fig:IVF}
\end{figure*}

The timeline of a typical IVF procedure is shown in Fig.~\ref{fig:IVF}. During the patient's first visit, initial consultation is conducted, her medical history is recorded, and basic medical examination is performed. This process may take 1-2 days. At Day~3, the patient's basic characteristics such as age, BMI, infertility duration, AFC, AMH, FSH, pathogenesis, etc., are available. If the patient determines to perform IVF, then usually it will take three menstrual cycles. In the first menstrual cycle, additional examination and controlled ovarian hyper-stimulation (COH) are performed. Oocyte pickup and egg fertilization are done in the second menstrual cycle. Embryo or balstocyst transfer are performed in the third menstrual cycle. The entire process takes about 2-3 months. During this process, embryo morphology features can be extracted to determine the embryo quality, number of embryo to transfer, and the transfer plan, etc. If the patient fails to conceive after embryo transfer, she has to spend the same amount of time again to repeat this procedure, which represents a heavy burden to many patients, economically, physically, and emotionally.

Cumulative pregnancy rate, which tells the probability that a patient conceives pregnancy after multiple IVF cycles, is an important measure for evaluating different IVF approaches, and is usually also the first question that a patient asks before starting the IVF. Given the long duration (2-3 months) and high cost of an IVF cycle (the average cost of an IVF cycle is approximately \$10,000-15,000 in the United States \cite{Hornstein2016}, and \$4,500 in Tongji Hospital in China), it is important to be able to accurately estimate the individualized cumulative pregnancy rate, so that the patient can make the most appropriate decisions on whether to use her own oocyte or donor oocyte, how many oocyte pickup cycles she may need, whether to use embryo frozen, etc. Artificial intelligent, particularly machine learning \cite{Bishop2006}, could be used for this purpose.

Machine learning has rapidly progressed the medical field during the past few years. It has been used to predict the development of hepatocellular carcinoma \cite{Singal2013}, adult autism spectrum disorder \cite{Yahata2016}, non-small cell lung cancer prognosis \cite{Yu2016}, human oocyte developmental potential \cite{Yanez2016}, the risk of acute myeloid leukaemia \cite{Abelson2018}, etc., and also to identify a human neonatal immune-metabolic network associated with bacterial infection \cite{Smith2014}, to classify skin cancer \cite{Esteva2017}, to isolate individual cell for scalable molecular genetic analysis of single cells \cite{Brasko2018}, and so on.

Machine learning has also been used to predict the pregnancy result with features obtained before and during the IVF, including basic patient characteristics, embryo morphology, and so on. For example, decision trees \cite{Quinlan1986,Quinlan1992} have been used to investigate the relationship between the outcome of transfer and 53 embryo, oocyte and follicular features \cite{Saith1998}, to predict the IVF outcome from 100 variables related to the basic patient characteristics (e.g., age, body mass index, etc.) and derived from the different stages of the IVF cycle (e.g., the amount of hormone treatment, the measurement of ovary volume, etc.) \cite{Passmore2003}, and to predict the IVF outcome from 69 features on patient's basic information, diagnosis, clinical tests, treatment methods, etc \cite{Guh2011}. Bayesian classifiers have been used to select the most promising embryos to transfer to the woman's uterus using features related to clinical data and embryo morphology \cite{Morales2008}, and to predict implantation outcome of individual embryos in an IVF cycle from 18 features including age, infertility factor, treatment protocol, sperm, embryo morphology, etc \cite{Uyar2015}. Support vector machines (SVMs) \cite{Uyar2008} and Bayesian Classifiers \cite{Uyar2009} have been used to predict implantation outcomes of new embryos from 17 features related to patient characteristics, clinical diagnosis treatment method, and embryo morphological parameters. However, to our knowledge, no one has used only patient characteristics from basic medical examinations to predict the cumulative IVF pregnancy rate, as we are doing in this study.

In this paper, we propose supervised and unsupervised machine learning approaches for cumulative pregnancy rate prediction from basic patient characteristics. We show that the approach that integrates unsupervised learning and supervised learning achieves the best performance. Our approach can significantly save the time and cost in predicting the cumulative IVF pregnancy rate, and thus can help the patients make more appropriate decisions before the IVF starts.

The remainder of this paper is organized as follows: Section~\ref{sect:method} introduces our three machine learning approaches for cumulative pregnancy rate prediction. Section~\ref{sect:results} presents the experimental results. Section~\ref{sect:discussion} discusses the benefits of our proposed approaches. Finally, Section~\ref{sect:conclusion} draws conclusion.

\section{Our Proposed Machine Learning Approaches} \label{sect:method}

This section introduces the dataset used in our study, and the feature selection and machine learning approaches for cumulative pregnancy rate prediction from basic patient characteristics.

\subsection{The Dataset}

This study consisted of 11,190 Chinese couples who suffered from infertility and received IVF treatments at Tongji Hospital (ranked 3rd in Gynaecology and Obstetrics in China), Huazhong University of Science and Technology, Wuhan, China, between January 2016 and March 2018. Their IVF cycles varied from one to 11, as summarized in Table~\ref{tab:summary}. Only basic patient characteristics obtained from the initial medical examination were used in our prediction, which included female age, female body mass index (BMI), infertility duration, antral follicle count (AFC), anti-mullerian hormone (AMH), follicle-stimulating hormone (FSH), and 30 pathogeny factors.

\begin{table}\centering
\caption{Summary of basic patient characteristics in our study. The first six features are numerical. Their means and standard deviations are calculated. Pathogeny has 30 factors. For each factor, the number of patients and the percentage are given. The 11 used features are marked by asterisks.} \label{tab:summary}
\begin{tabular}{ll}\hline
\textbf{Cycle Statistics}       &           \\ \hline
Cycle 1         & n=9,419    \\
Cycle 2         & n=1,432    \\
Cycle 3         & n=236      \\
Cycle 4         & n=59       \\
Cycle 5         & n=30       \\
Cycle 6         & n=7        \\
 Cycle 7        & n=2        \\
 Cycle 8        & n=2        \\
Cycle 9         & n=2        \\
Cycle 10        & n=0        \\
Cycle 11        & n=1        \\
 Total          & n=11,190    \\ \hline
\textbf{Patient Characteristics}      &     \\ \hline
Female age  (years)*                    & 31.5 $\pm$ 5.22     \\
Female BMI  ($kg/m^2$)*                 & 21.85 $\pm$ 2.90    \\
Infertility duration  (years)*          & 3.64 $\pm$ 2.98     \\
Antral follicle count (AFC)*         & 12.93 $\pm$ 7.08    \\
Anti-mullerian hormone (AMH)  ($ng/ml$)*      & 4.95 $\pm$ 4.07     \\
Follicle-stimulating hormone (FSH) ($IU/L$)* & 7.94 $\pm$ 3.12     \\
Pathogeny:     &  \\
\quad Ovulatory dysfunction                 & n=53  (0.5\%)        \\
\quad Polycystic ovary syndrome (PCOS)*      & n=1,123  (10.0\%)     \\
\quad Abnormal uterine bleeding  (AUB)       & n=1  (0.0\%)         \\
\quad Hypogonadolropic hypogonadism  (HH)    & n=17  (0.2\%)        \\
\quad Kallmann syndrome                     & n=0  (0.0\%)         \\
\quad Hyperprolactinemia                    & n=55  (0.5\%)        \\
\quad Pituitary adenoma                     & n=16  (0.1\%)        \\
\quad Panhypopituitarism                    & n=0  (0.0\%)         \\
\quad Empty sella syndrome  (ESS)            & n=0  (0.0\%)         \\
\quad Diminished ovarian reserve  (DOR)*      & n=1,512  (13.5\%)     \\
\quad Premature ovarian insufficiency  (POI) & n=2  (0.0\%)         \\
\quad Perimenopause*                         & n=641  (5.7\%)       \\
\quad Pelvic inflammatory disease  (PID)     & n=1,326  (11.8\%)     \\
\quad Tubal obstruction                     & n=2,451  (21.9\%)     \\
\quad Hydrosalpinx                          & n=388  (3.5\%)       \\
\quad Salpingitis                           & n=2,306  (20.6\%)     \\
\quad Pelvic tuberculosis                   & n=48  (0.4\%)        \\
\quad Endometriosis                         & n=494  (4.4\%)       \\
\quad Chocolate cyst                        & n=372  (3.3\%)       \\
\quad Adenomyoma                            & n=239  (2.1\%)       \\
\quad Uterine malformation                  & n=182  (1.6\%)       \\
\quad Intrauterine adhesion*                 & n=417  (3.7\%)       \\
\quad Scarred uterus                        & n=857  (7.6\%)       \\
\quad Myoma                                 & n=460  (4.1\%)       \\
\quad Endometritis                          & n=20  (0.1\%)        \\
\quad Endometrial tuberculosis              & n=21  (0.1\%)        \\
\quad Endometrial hyperplasia  (EH)          & n=4  (0.0\%)         \\
\quad Chromosome abnormality                & n=208  (1.9\%)       \\
\quad Paternal factor*                       & n=3,253  (29.1\%)     \\
\quad Others                       & n=37 (0.3\%)     \\ \hline
\end{tabular}
\end{table}

\subsection{Feature Selection}

In order to select the most informative features, we performed logistic regression \cite{HosmerJr2013} using all basic patient characteristics, where each categorical feature was converted to a binary value using one-hot encoding. We used only Cycle~1 pregnancy results as the labels for logistic regression, and excluded patients who did not receive a transfer in Cycle~1.

Multiple logistic regression analyses showed that 14 features had significant correlation with pregnancy results ($P < 0.01$). Among them, three etiological factors (endometrial tuberculosis, chromosome abnormality, and others) had fewer than 2\% of the total patients. They were removed to make the features more representative. As a result, 11 features were finally selected for further analysis, and they are marked by asterisks in Table~\ref{tab:summary}.

\subsection{Cumulative Pregnancy Rate Prediction} \label{sec:methods_brief}

The prediction of IVF outcome is extremely difficult using only basic patient characteristics without controlled ovarian hyper-stimulation details, and embryo and endometrial features. According to previous research, embryo features are very important for the final outcome prediction using machine learning \cite{Guh2011, Milewski2011}. When using only basic patient characteristics, we assume that patients having similar basic characteristics also have similar pregnancy rates. This is the best assumption we could make before starting the actual IVF. When the patients start the IVF, more features could be extracted, and more individualized prediction could be made. However, these features are not available before the IVF, and hence will not be used in our model.

\begin{figure*} \centering
\includegraphics[width=.7\linewidth]{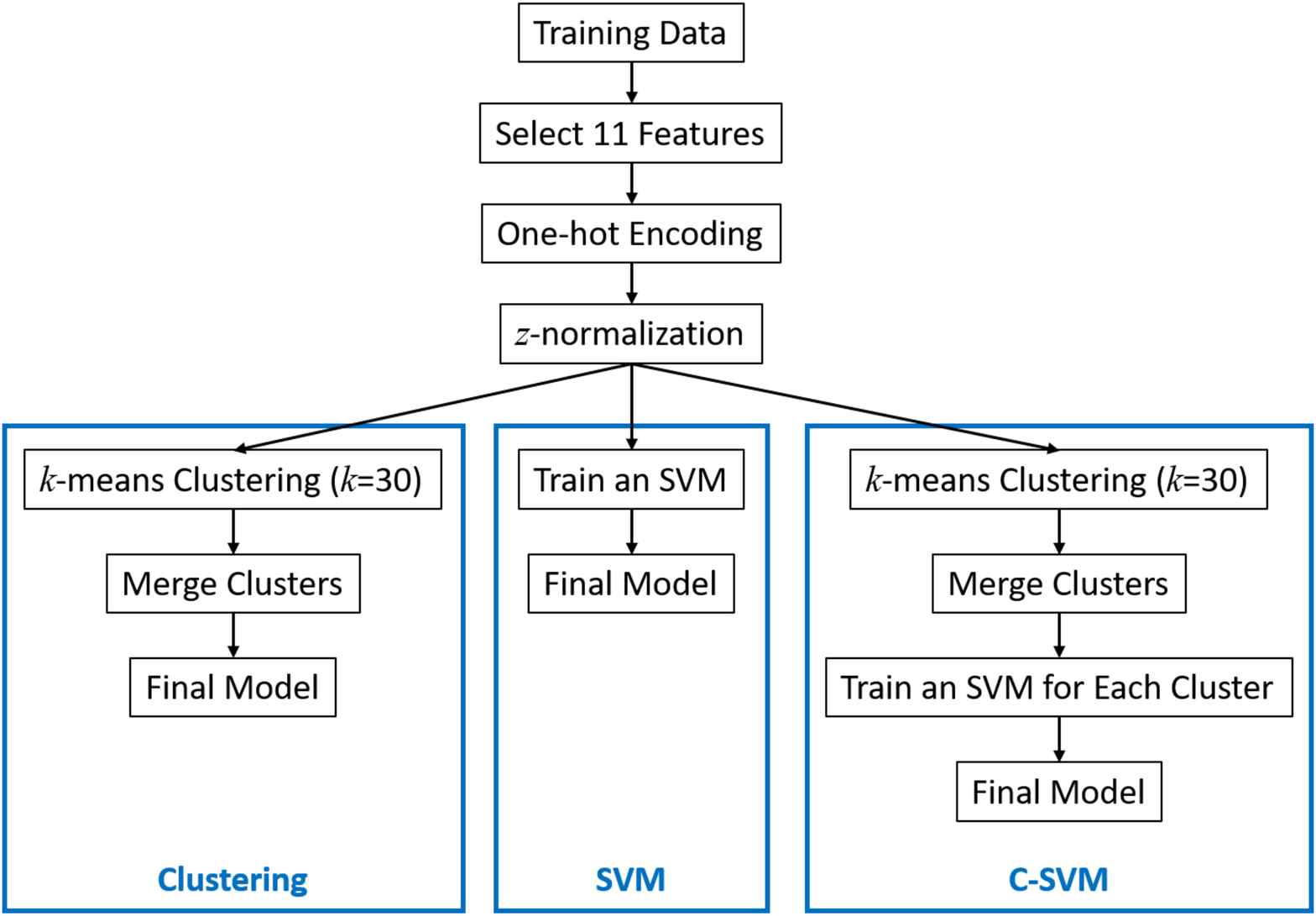}
\caption{Pipeline of the three proposed machine learning approaches. Given a training dataset of patients with basic medical examination information, all three models use the 11 asterisk features in Table~\ref{tab:summary}, convert the categorical features to numerical features using one-hot encoding, and $z$-normalize each feature. Clustering is an unsupervised approach. SVM is a supervised approach. C-SVM integrates both unsupervised and supervised approaches.} \label{pic:method}
\end{figure*}

We constructed three different machine learning models -- clustering, SVM, and clustering-SVM (C-SVM), and compared their performances using three measures. The pipeline of our three machine learning approaches is shown in Fig.~\ref{pic:method}. Only the 11 asterisk features in Table~\ref{tab:summary} were used. We first used one-hot encoding to convert each categorical feature into numerical features, and then performed $z$-normalization to transform each feature to have mean 0 and standard deviation 1.

\subsection{Model~1: Clustering}

In the training phase of the clustering approach, we first applied $k$-means clustering with $k=30$ to all patients. We then identified all possible $30\times 29/2=435$ unique pairs of clusters. For each pair, we performed the log-rank test \cite{verwaal2003randomized, goel2010understanding,tan1992cumulative} between the two clusters to check if the difference between them was significant. If the $p$ value of at least one of the 435 tests was larger than a predefined threshold $\alpha$ ($\alpha=0.01$ was used in our study), then we identified the two clusters with the largest $p$-value (which meant the two clusters were the most similar) and merged them. We repeated the log-rank tests with the remaining clusters, until all $p$-values were smaller than $\alpha$. We then recorded the center of each cluster, and its corresponding cumulative pregnancy rate.

In the testing phase, when the basic characteristics of a new patient came in, we assigned the patient to the cluster with the closest centroid, and then used the corresponding cumulative pregnancy rate as the prediction.

\subsection{Model~2: SVM}

For the SVM classifier \cite{Vapnik1998}, we first performed 5-fold cross validation on the training set to search for the best kernel function (polynomial, RBF, or linear) and to determine whether a larger weight should be used to accommodate the minority class. Eventually we used the RBF kernel and set the per-class weights inversely proportional to class frequencies in the training data. We then used penalty parameter $C=1$ to train a probabilistic SVM classifier.

\subsection{Model~3: C-SVM}

The C-SVM approach was a sequential combination of the clustering approach and the SVM approach. In the training phase, it first used the clustering approach to group the patients into several clusters, and then trained an RBF SVM for each cluster to individualize the patients within each cluster.

In the testing phase, when the basic characteristics of a new patient came in, we first assigned the patient to the cluster with the closest centroid, and then used the corresponding SVM to predict a more individualized cumulative pregnancy rate.

\section{Prediction Results} \label{sect:results}

This section compares the prediction performances of the three proposed approaches.

\subsection{Area under the Curve (AUC)}

\begin{figure} \centering
\includegraphics[width=\linewidth]{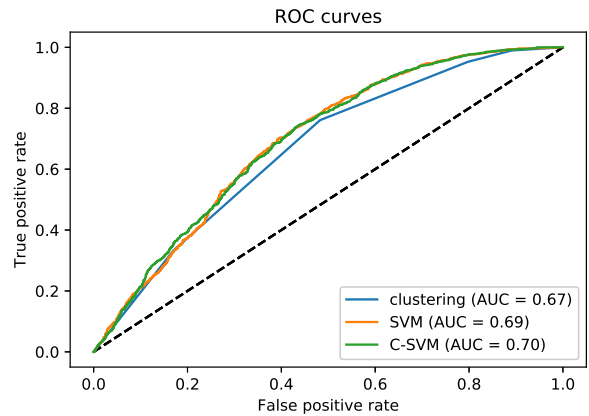}
\caption{ROC curves and AUCs of the three approaches.} \label{fig:ROC}
\end{figure}

First, we evaluated the performances of the three approaches by randomly sampling two thirds of the patients as training data, and the remaining one third as test data. We used the training data to train the three models and then validated them on the test data. Their receiver operating characteristic (ROC) curves are shown in Fig.~\ref{fig:ROC}, and the corresponding areas under the curve (AUCs) were also computed and indicated in the legend. Fig.~\ref{fig:ROC} shows that SVM and C-SVM had similar AUC performances (0.69 and 0.70, respectively), both of which were higher than clustering (AUC=0.67).

\subsection{Cumulative Pregnancy Rate Prediction}

\begin{figure}[htpb]\centering
\subfigure[]{\label{fig:CPR1}   \includegraphics[width=\linewidth,clip]{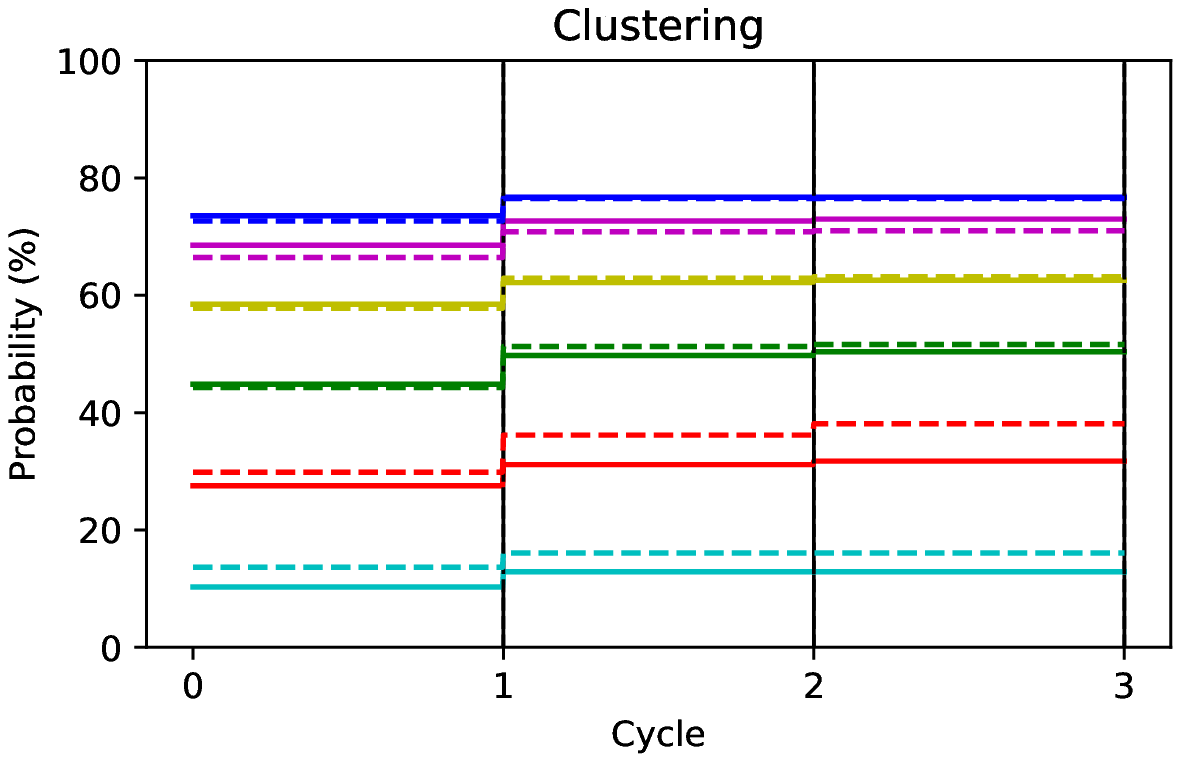}}
\subfigure[]{\label{fig:CPR2}   \includegraphics[width=\linewidth,clip]{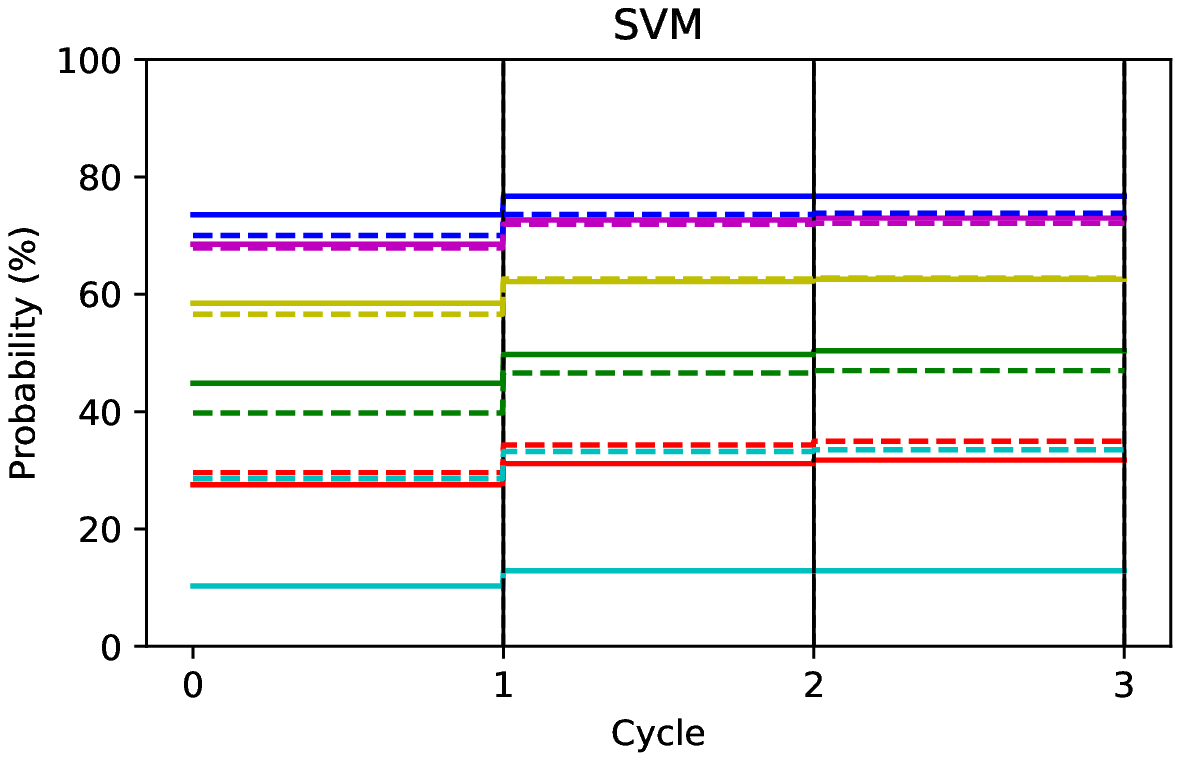}}
\subfigure[]{\label{fig:CPR3}   \includegraphics[width=\linewidth,clip]{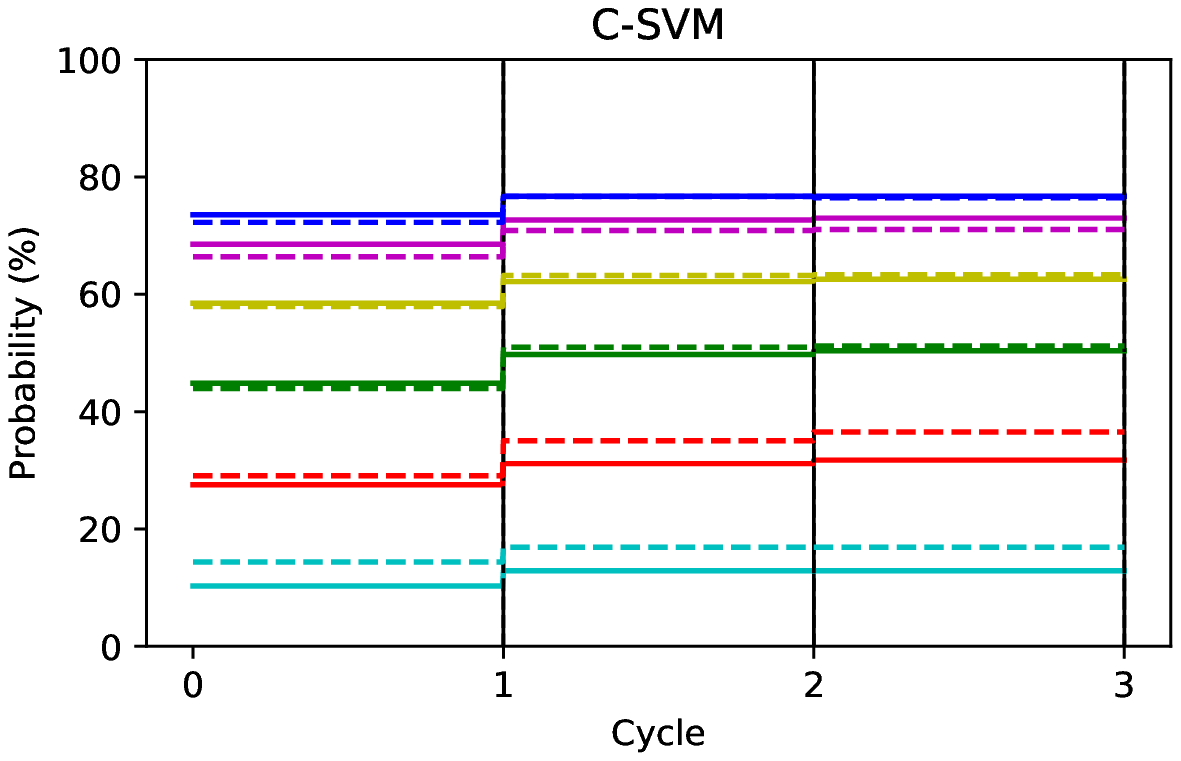}}
\caption{Predicted cumulative pregnancy rate curves of the three models on test data. (a) Clustering; (b) SVM; (c) C-SVM. The solid curves in the same color in the three subfigures are identical, indicating the true cumulative pregnancy rate curve of a cluster. The dashed curves indicate the predictions from different approaches. In each subfigure, the solid and dashed curves in the same color indicate results in the same cluster, which should be close.} \label{fig:CPR}
\end{figure}

Once we get the predicted probability and the corresponding cluster of each patient in test data, we can predict the cumulative pregnancy rate using the mean probability of the corresponding cluster. Fig.~\ref{fig:CPR} shows the cumulative pregnancy rate curve using the three approaches. Although SVM had promising AUC in Fig.~\ref{fig:ROC}, its cumulative pregnancy rate prediction had large biases. On the other hand, clustering and C-SVM, particularly C-SVM, had much smaller prediction errors.

\subsection{Stability of the Prediction Models}

In order to test the stability of the three prediction models, we repeated them 30 times, each time with different training and test data. As Table~\ref{tab:summary} shows that less than 1\% patients had more than three cycles, we did not consider cycle numbers larger than three.  The mean and standard deviation of the AUCs from the 30 runs are shown in the first part of Table~\ref{tab:rmseAUC}. On average C-SVM achieved the best AUCs in the three cycles.

\begin{table}[htpb] \centering \setlength{\tabcolsep}{2mm}
\caption{Mean and standard deviation (in parentheses) of AUCs and RMSEs of the three proposed approaches. The best ones are marked in bold.}
\label{tab:rmseAUC}
\begin{tabular}{c|c|ccc} \cline{1-5}
                           & Cycle & Clustering        & SVM                     & C-SVM            \\ \hline
\multirow{3}{*}{AUC} & 1     & 0.6725 (0.0075) & 0.6906 (0.0080)          & \textbf{0.6920 (0.0075)} \\
                           & 2     & 0.6699 (0.0065) & \textbf{0.6891 (0.0074)} & 0.6890 (0.0069)          \\
                           & 3     & 0.6684 (0.0063) & 0.6865 (0.0075)          & \textbf{0.6866 (0.0076)} \\\hline
RMSE & All   & 0.0274 (0.0098) & 0.0794 (0.0130)          & \textbf{0.0267 (0.0096)} \\ \hline
\end{tabular}
\end{table}

We also studied the stability of the three approaches using another the root mean squared error (RMSE). For each model in each run, we concatenated the predicted cumulative pregnancy rates in three cycles and $n$ clusters into a $3n$-element vector $\hat{\mathbf{y}}=[\hat{y}_1,...,\hat{y}_{3n}]$, and computed the RMSE between the predictions and the corresponding groundtruth $\mathbf{y}=[y_1,...,y_{3n}]$,
\begin{align}
RMSE = \left[\frac{1}{3n}\sum_{i=1}^{3n} (\hat{y}_{i}-y_{i})^{2}\right]^{1/2}
\end{align}
A smaller RMSE means a better performance. The mean and std of the RMSEs in the 30 runs are shown in the second part of Table~\ref{tab:rmseAUC}. Again, C-SVM achieved the best performance.

We also performed an analysis of variance (ANOVA) test to check if there was statistically significant difference between each pair of algorithms. The $p$-values are shown in Table~\ref{tab:anova}, where the statistically significant ones are marked in bold. SVM and C-SVM were statistically significantly better than clustering on AUC, and SVM and C-SVM were statistically significantly better than clustering on RMSE. In summary, C-SVM achieved the best overall performance.

\begin{table}[htpb]\centering \setlength{\tabcolsep}{5mm}
\caption{$p$-values of ANOVA tests ($\alpha=0.05$) on the three proposed approaches.} \label{tab:anova}
\begin{tabular}{c|c|cc} \hline
                      &             & Clustering       & SVM           \\ \hline
\multirow{2}{*}{AUC}  & SVM         & \textbf{0.00} &               \\
                      & C-SVM & \textbf{0.00} & 0.92          \\ \hline
\multirow{2}{*}{RMSE} & SVM         & \textbf{0.00} &               \\
                      & C-SVM & 0.97          & \textbf{0.00} \\ \hline
\end{tabular}
\end{table}

\section{Discussions} \label{sect:discussion}

This section discusses the advantages of our proposed approaches, particulary C-SVM, our best-performing model.

\subsection{C-SVM Reduce the Time and Cost to Predict the IVF Cumulative Pregnancy Rate}

Our C-SVM model uses only the basic medical examination information during the first visit (which takes two days and costs about \$50 in Tongji Hospital in China) to predict the cumulative pregnancy rate, and the result can be known immediately after the visit.

Compared with the conventional approaches in the literature, which use information during the IVF (which takes 2-3 months and costs about \$4,500 in Tongji Hospital in China), our approach is much faster and more economic. It significantly saves the patient's time and cost, and represents a step towards precision medicine and individualized treatment.

\subsection{Cumulative Pregnancy Rate Prediction Is More Informative than Single-Cycle Pregnancy Rate Prediction}

The total duration and cost of IVF is significantly impacted by the number of oocyte pickup cycles. Since oocyte pickup is time-consuming and expensive, a patient may choose to freeze the extra embryos from the first oocyte pickup cycle to reduce the time and cost: the frozen embryos can be transferred in case previous transfers fail, without the need to pickup fresh oocyte and fertilize them again. However, frozen embryo transfer may have a lower pregnancy rate than fresh embryo transfer. So, it is important to know the cumulative pregnancy rate of fresh embryo transfers so that the patient can make a smarter decision on whether it is worthwhile to save the time and cost of another oocyte pickup cycle. Our C-SVM model can predict the cumulative pregnancy rates after one, two or three oocyte pickup cycles, which gives the patients exact information they need in decision making.

A patient with poor ovarian reserve is very difficult to conceive using her own oocyte. Knowing the cumulative pregnancy rate using her own oocyte could greatly help her make a wiser decision: if the cumulative pregnancy rate using her own oocyte is much lower than her expectation, then the patient may choose to receive donor oocyte, which may have much higher pregnancy rate. In this way, the patient can avoid potentially multiple controlled ovarian hyper-stimulations, shorten the time to pregnancy, reduce the overall cost, and hence improve the quality of life.

\section{Conclusion} \label{sect:conclusion}

In this paper, we have developed three different approaches (clustering, SVM, and C-SVM) to predict the IVF cumulative pregnancy rate in multiple cycles of oocyte pickup using basic patient characteristic. The selected parameters included female age, female BMI, infertility duration, AFC, AMH, FSH, and five pathogeny factors (diminished ovarian reserve, perimenopause, paternal factor, PCOS, and intrauterine adhesion). Experimental results showed that the AUCs of SVM and C-SVM were better than that of clustering, and the prediction RMSEs of clustering and C-SVM were smaller than that of SVM. In summary, C-SVM seems to be the best model.

To our best knowledge, this is the first study on using machine learning to predict the cumulative pregnancy rate of multiple IVF cycles from only basic patient characteristics before the actual IVF. The predictions can help the patient make optimal decisions on whether to use her own oocyte or donor oocyte, how many oocyte pickup cycles she may need, whether to use embryo frozen, etc. They will also reduce the patient's cost and time to pregnancy, and improve her quality of life.


\end{document}